\documentclass[aps,amsmath,preprint,showpacs]{revtex4}
\usepackage{graphicx}
\usepackage{dcolumn}
\usepackage{color}

\newcommand{\tauiso}{{\mbox{\boldmath $\tau$}}}

\newcommand{\vecvarphi}{{\mbox{\boldmath $\varphi$}}}

\newcommand{\bm}{\bibitem}

\newcommand*{\s}[1]{/\llap{$#1$}} 
\newcommand*{\lrpart}{\tensor\partial}
\newcommand*{\dop}[2]{#1\cdot#2}

\newcommand*{\res}[2]{\ensuremath{#1_{#2}}}

\newcommand*{\gamF}{\gamma_5}

\newcommand*{\eps}{\varepsilon}

\newcommand*{\A}{\ensuremath{\mathsf A}}

\newcommand*{\K}{\ensuremath{\mathsf K}}
\newcommand*{\T}{\ensuremath{\mathsf T}}
\newcommand*{\tf}{\ensuremath{\tilde f}}

\newcommand*{\La}{\ensuremath{\mathcal L}}

\newcommand*{\oh}{\ensuremath{\frac12}}

\newcommand*{\mc}[1]{\multicolumn{1}{c|}{#1}}
\newcommand*{\mcc}[1]{\multicolumn{1}{c}{#1}}
\newcommand*{\N}{\mc{---}}  
\newcommand*{\Nc}{\mcc{---}}  

\newcommand*{\ch}[4]{\ensuremath{#1 #2 \to #3 #4}}
\newcommand*{\YYv}[1]
    {\frac{(\chi #1 + i \s\partial #1 /2m_N) \cdot \tauiso}
         {\chi+1} \gamF}
\newcommand*{\YY}[1]
    {\frac{\chi #1 + i \s\partial #1 /2m_N}
         {\chi+1} \gamF}
\newcommand*{\XXv}[1]
    {\left(\gamma_\mu \vecvarphi_\rho^\mu
    +\frac{\kappa_{\rho}}{2m_N} \sigma_{\mu\nu}\partial^\nu\vecvarphi_\rho^\mu
      \right) \cdot \tauiso}
\newcommand*{\XX}[1]
    {\left(\gamma_\mu \varphi_\omega^\mu
    +\frac{\kappa_\omega}{2m_N} \sigma_{\mu\nu}\partial^\nu \varphi_\omega^\mu
      \right)}
\newcommand*{\XXp}[1]
    {\left(\gamma_\mu \varphi_\phi^\mu
    +\frac{\kappa_\phi}{2m_N} \sigma_{\mu\nu}\partial^\nu \varphi_\phi^\mu
      \right)}
\newcommand*{\EPS}[4]
    {\left(\eps_{\mu\nu\rho\sigma}
      (#1^\rho #2^\mu) (#3^\sigma #4^\nu) \right)}
\long\def\Omit#1{}

\newcolumntype{x}[1]{D..{#1}}
\newcolumntype{C}{>{$}c<{$}}
\newcolumntype{R}{>{$}r<{$}}

\newcommand*{\tblref}[1]{Table~\ref{tbl:#1}}
\newcommand*{\tbllab}[1]{\label{tbl:#1}}
\renewcommand*{\eqref}[1]{Eq.~(\ref{eq:#1})}
\newcommand*{\eqlab}[1]{\label{eq:#1}}

\newcommand*{\secref}[1]{Section~\ref{sec:#1}}
\newcommand*{\seclab}[1]{\label{sec:#1}}

\begin{document}

\title{Photoproduction of $\eta$ meson within a coupled-channels \K-matrix
approach}
\author{R. Shyam$^1$ and O. Scholten$^2$}
\affiliation{$^1$Saha Institute of Nuclear Physics, Kolkata 70064, India\\
$^2$Kernfysisch Versneller Instituut, University of Groningen, 9747 AA,
Groningen, The Netherlands }
\date{\today}

\begin{abstract}

We investigate photoproduction of $\eta$ mesons off protons and
neutrons within a coupled-channels effective-Lagrangian method which is
based on the \K-matrix approach. The two-body final channels included are
$\pi N$, $\eta N$, $\phi N$, $\rho N$, $\gamma N$, $K \Lambda$, and
$K \Sigma$. Non-resonant meson-baryon interactions are included in the model
via nucleon intermediate states in the $s$- and $u$-channels and
meson exchanges in the $t$-channel amplitude and the $u$-channel resonances.
The nucleon resonances $S_{11}$(1535), $S_{11}$(1650), $S_{31}$(1620),
$P_{11}$(1440), $P_{11}$(1710), P$_{13}$(1720), $P_{33}$(1232),
$P_{33}$(1600), D$_{13}$(1520), $D_{13}$(1700), and $D_{33}$(1700) are
included explicitly in calculations. Our  model describes simultaneously
the available data as well on total and differential cross sections as on
beam and target asymmetries. This holds for the $\gamma p \to \eta p$ reaction
for photon energies ranging from very close to threshold to up to 3~GeV.
The polarization observables show strong sensitivity to resonances that
otherwise contribute only weakly to the total cross section. It is found
that the pronounced bump-like structure seen in the excitation function of
the $\gamma n \to \eta n$ cross section at $\gamma$ energies around 1~GeV, can
be explained by the interference effects of $S_{11}$, $P_{11}$ and $P_{13}$
resonance contributions.

\end{abstract}
\maketitle

\newpage
\section{Introduction}

It is well established that nucleons have a rich excitation spectrum which
reflects their complicated multi-quark inner dynamics. The determination of
properties of the nucleon resonances (e.g., their masses, widths, and coupling
constants to various decay channels) is an important issue in hadron physics.
This will provide the benchmark for testing the predictions of lattice
quantum chromodynamics (LQCD) which is the only theory which tries to
calculate these properties from first principles~\cite{wil74}. Even though,
the requirement of computational power is enormous for their numerical
realization, such calculations have started to provide results for nucleon
properties for its ground as well as excited states~\cite{bur06,lei05}.
Furthermore, reliable nucleon resonance data are also important for testing the
"quantum chromodynamics (QCD) based" quark models of the nucleon (see,
{\it e.g.},~\cite{cap00,lor01}) and also the dynamical coupled-channels models
of baryonic resonances~\cite{lut05}.

Experimental determination of baryonic resonance properties proceeds indirectly
by exciting the nucleon with the help of a hadronic or electromagnetic probe
and performing measurements of their decay products (mesons and nucleons). The
reliable extraction of nucleon resonance properties from such experiments is a
major challenge. Description of intermediate-energy scattering is still too
far away from the scope of the LQCD calculations.  Therefore, at this stage the
prevailing practice is to use effective methods to describe the dynamics of
meson production reactions. Such methods include explicit baryon resonance
states, whose properties are extracted by comparing the prediction of the
theory with the experimental data~\cite{ben95,arn00,wen02,tia99,dre99,shy07}.

In order to determine resonance properties reliably from the experimental
measurements one requires a model that can analyze the different reactions
over the entire energy range using a single Lagrangian density that
generates all non-resonance contributions from Born, $u$- and $t$-channel
contributions without introducing new parameters. At the same time, the
Lagrangian should also satisfy the symmetries of the fundamental theory (i.e.
QCD) while retaining only mesons and baryons as effective degrees of freedom.
Conformity with chiral symmetry  is known to be important for low
energy pion-nucleon physics.

A way to analyze simultaneously all reaction data for a multitude of
observables in different reaction channels while respecting the constrains
described above, is provided by a coupled-channels method within the \K-matrix
approach~\cite{feu98,kor98,pen02,uso05,shk05,uso06}. This method is attractive
because it is based on an effective-Lagrangian framework that is gauge
invariant and is consistent with chiral symmetry. It also provides a convenient
way of imposing the unitarity constraint. This results from the Bethe-Saltpeter
equation in the approximation where only the discontinuity part of the loop 
integral is retained i.e, the particles forming the loop are taken on the mass
shell. The $S$ matrix in this approach is unitary provided the \K-matrix is 
taken to be real and Hermitian.

Alternatively, the dynamical coupled channels models within the Hamiltonian
formalism have also been used to describe the meson-production reactions
\cite{dur08,jul08,jul07}. A relativistic chiral unitary approach based on a
coupled-channels prescription has been used in Ref.~\cite{bor02} to calculate
cross sections for $\eta$ photoproduction. This  reaction has also been studied
in chiral perturbation theory~\cite{bor00} and in the chiral constituent-quark
model~\cite{sag01}.

In this paper we investigate photoproduction of the $\eta$ meson
which is the next lightest non-(open)strange member in the meson mass
spectrum. This is a subject of considerable interest. It can be used to
probe the $s{\bar s}$ component in the nucleon wave function~\cite{dov90}.
There is also interest in measuring the rare decays of $\eta$ mesons which
could provide a new rigorous test of the standard model~\cite{pap05} or even
of the physics beyond this. The nucleon resonance $N^*$(1535) [$S_{11}(1535)$]
with spin-$\frac{1}{2}$, isospin-$\frac{1}{2}$, and odd parity, has a
remarkably large $\eta N$ branching ratio. It lies only about 50~MeV above the
$N\eta$ production threshold, and contributes dominantly to the photoproduction
amplitude at energies close to threshold. Thus this reaction is an ideal
tool to study the $N^*$(1535) resonance which has been the subject of some
debate recently (see, e.g.,~\cite{mat05}). The attractive nature of the
$\eta$-nucleon interaction may lead to the formation of bound (quasi-bound)
$\eta$-nucleus states (see, e.g.,~\cite{chi91,fix02,hai02,pfe04,sib04}).

Photoproduction reactions provide a sensitive tool to study baryonic
resonances. Apart from giving information which is complementary to
that extracted from the studies of hadronic reactions, it gives access to
additional information about the weakly excited resonances through the
polarization observables~\cite{kno95}. The $\eta$ meson has zero isospin
($I$), hence the $\eta N$ final states can only be reached via excitation
of $I = \frac{1}{2}$ resonances ($N^*$). This is in contrast to the $\pi N$
channel where both $I = \frac{1}{2}$ and $I = \frac{3}{2}$ intermediate
states are possible. Thus even if a resonance has only a small coupling to
the $\eta N$ channel, it is identified as a N$^*$ state.

With the advent of new high-duty-cycle electron accelerators and intense photon
sources with sophisticated detectors, a rich variety of very precise data
have been accumulated on total and differential cross sections and beam and
target asymmetries for $\eta$ meson photoproduction off the free proton
\cite{pri95,kru95,aja98,boc98,tho01,dug02,ren02,cre05,bar07,els07,bar07a}.
For a comprehensive review of the data up to 2003 and their interpretations,
we refer to~\cite{kru03}. The eta-Maid~\cite{wen02} and partial wave
analyses~\cite{ani05} of these data reveal that while the cross sections are
dominated by the excitation of the S$_{11}$(1535) resonance in the threshold
region and by the P$_{13}$(1720) resonance at higher photon energies, beam
and target asymmetries are sensitive to the weakly excited D$_{13}$(1520)
and P$_{11}$(1710) resonances via interference with the strongly excited ones.

Extensive measurements have also been performed for the $\eta$ meson production
on a deuterium target~\cite{kru95,hof97,wei01,wei03,kuz07,jae08}. These data
have provided useful information on the isospin structure of electromagnetic
excitations of the $S_{11}$(1535) resonance and have led to the determination
of the $\gamma n \to \eta n$ / $\gamma p \to \eta p$ cross-section ratio.  An
interesting observation of the data on the (quasi-free) neutron is that the
corresponding excitation function of the total cross section shows a bump-like
structure around photon energies of 1~GeV - such a bump is not seen in
the proton case. This bump structure has been explained either in terms of the
presence of a D$_{15}$(1675) resonance with an unusually large branching ratio
for its decay to the $\eta N$ channel~\cite{wen02} or due to coupled channels
effects involving S$_{11}$(1535), S$_{11}$(1650) and P$_{11}$(1710) resonances
\cite{shk07} within a coupled-channels effective Lagrangian model. In the
latter approach, contributions of spin-$\frac{5}{2}$ resonances are found to be
negligibly small. In an yet another explanation, it has been suggested that
this bump may be a signal of the existence of a relatively narrow (width $<$
30~MeV) baryonic state with mass around 1.68~GeV~\cite{kuz07,fix07}.

The main objective of this paper is to study photoproduction of $\eta$
mesons on the bare proton and neutron for photon energies ranging from 
threshold to 3~GeV in a coupled-channels formalism which is based on the
\K-matrix approach. This is an effective Lagrangian model which is gauge 
invariant and obeys the low-energy theorem. We aim at describing simultaneously
the data on total and differential cross sections and beam and target 
asymmetries. As described above the data base on $\eta$ photoproduction has 
been enhanced appreciably during recent times. It is a challenge to any 
theoretical model to describe all the available data within one framework.

We would like to add that our work, in a way, supplements the coupled-channels
effective-Lagrangian model calculations of this reaction presented within the
Giessen model~\cite{shy07}. Although the two approaches are similar in physics
contents, they differ in some details. An important difference lies in the
inclusion of reaction channels leading to states outside the model space.  
The Giessen model parameterizes the $2\pi N$ final state, as enters for 
example in the $\gamma N \to 2\pi N$ reaction, by an effective $\xi N$ state, 
where $\xi$ is an isovector scalar meson with mass $m_\xi = 2m_\pi$. In our 
model the coupling to states outside the model space is taken into account by
allowing for an energy dependent decay width to these states in the 
propagators for the different resonances. Another important difference lies in
the choice of contact terms and form factors. As shown in Ref~\cite{uso05} 
contact terms in photo-induced reactions, which are magnetic in origin, are 
not important at low energies but have a considerable effect at higher 
energies. A more minor difference is that in the present calculation we have 
chosen to exclude spin-$\frac{5}{2}$ resonances from our study as their 
contributions are shown~\cite{shk07} to be almost negligible to $\eta$ 
photoproduction in comparison to those of the dominant lower-spin states. 
Unlike the Giessen work we have not performed a full blown $\chi^2$ fitting of
all the available $\gamma N \to \eta N$ data. Rather, we have used the 
parameter set obtained in a previous work within our model~\cite{uso05,uso06} 
and have made adjustments in some of them so as to describe the $\eta N$ 
channel.

Our paper is organized in the following way. An overview of our model is given
in section II. This consists of a short discussion of the \K-matrix
formalism, the model space and the channels included, the Lagrangians and
the form factors. Our results and a discussion thereof are presented in
section III. Summary and conclusions of our work is presented in section
IV.

\section{Description of the Model}

This work is based on an effective-Lagrangian model.  The kernel in the
\K-matrix approach is built by using the effective Lagrangian which is
given in Appendix A. We have taken into account contributions from (i) the
nucleon Born term, (ii) $t$-channel exchanges of mesons, (iii) nucleon and
resonance terms in the $u$-channel, and (iv) baryonic resonance in the
$s$-channel (see Fig.~1). The sum of amplitudes (i), (ii) and (iii) is
termed as the background contribution in the following. As is discussed
below, this approach allows to account for coupled-channels effects while
preserving many symmetries of a full field-theoretical method.
\begin{figure*}
\begin{center}
\includegraphics[width=0.5 \textwidth]{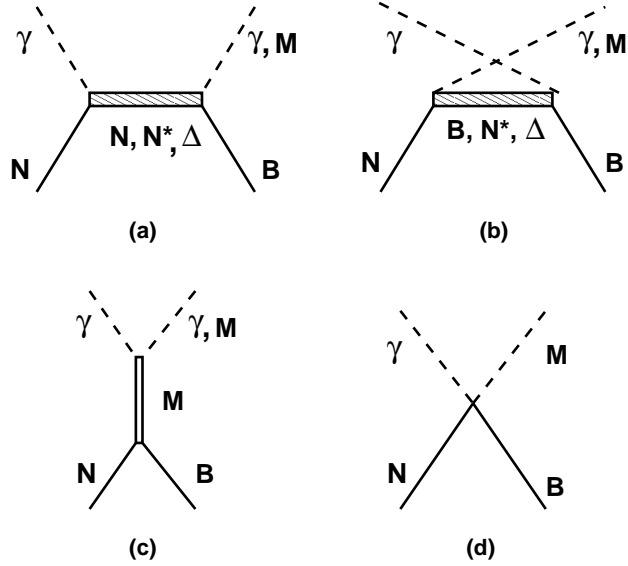}
\vskip -0.1in
\caption{[color online]
Feynman diagrams included in this work. First row: $s$- and $u$-channel
diagrams with propagating final state baryons ($B = N$, $\Lambda$, $\Sigma$)
or intermediate state resonances ($\Delta$, $N^*$). $M$ stands for the
mesons included in the model space. Second row: $t$-channel contributions
with propagating asymptotic and intermediate mesons, and the contact term
required by the gauge invariance.
}
\end{center}
\end{figure*}

\subsection{\K-matrix model}\label{sec:Kmatr}

The coupled-channels (or re-scattering) effects are included in our model
via the \K-matrix formalism. In this section we present a short overview of
this approach; a more detailed description can be found in Refs.
\cite{kor98,uso05,sch02,new82}.

In the \K-matrix formalism the scattering matrix is written as
\begin{equation}\eqlab{T-matr}
  \T = \frac{\K}{1-i\K} \,.
\end{equation}

It is easy to check, that the resulting scattering amplitude $S=1+2i\T$ is
unitary provided that \K\ is Hermitian. The construction in \eqref{T-matr} can
be regarded as the re-summation of an infinite series of loop diagrams by
making a series expansion,
\begin{equation}
  \T = \K + i\K\K + i^2\K\K\K + \cdots \,.
\end{equation}
The product of two \K-matrices can be rewritten as a sum of different one-loop
contributions (three- and four-point vertex and self-energy corrections)
depending on the Feynman diagrams that are included in the kernel \K.
However, not the entire spectrum of loop corrections present in a true
field-theoretical approach, is generated in this way and the missing ones
should be accounted for in the kernel. In constructing the kernel, care
should be taken to avoid double counting. For this reason we include in
the kernel tree-level diagrams only [Figs 1(a)-1(c)], modified with
form-factors and contact terms [Fig. 1(d)]. The contact terms (or four-point
vertices) ensure gauge invariance of the model and express model-dependence
in working with form factors (see~\secref{ff}). Contact terms and form factors
can be regarded as accounting for loop corrections which are not generated in
the \K-matrix procedure, or for short-range effects which have been omitted
from the interaction Lagrangian. Inclusion of both $s$- and $u$-channel
diagrams [Figs 1(a) and 1(b), respectively] in the kernel insures the
compliance with crossing symmetry.

{\squeezetable
\begin{table}
  \caption{ \tbllab{resonances}
    Baryon states included in the calculation of the
    kernel with their coupling constants. The column labeled
    WD lists the decay width to states outside the model space.
    The columns labeled M and WD are in units of GeV. See text
    for a discussion on the signs of the coupling constants.}
  \begin{ruledtabular}
  \begin{tabular}{C|x3|x3|d|d|d|d|d|d|d|d} 
    L_{IJ} &\mc{M}&\mc{WD}&\mc{$g_{N\pi}$} &\mc{$g^1_{p\gamma}$}
    &\mc{$g^2_{p\gamma}$}&\mc{$g^1_{n\gamma}$}&\mc{$g^2_{n\gamma}$}
             &\mc{$g_{K\Lambda}$}&\mc{$g_{K\Sigma}$}&\mcc{$g_{N\eta}$} \\
    \hline
    N           &0.939 &0.0  &13.47&\N   & \N & \N  & \N &12    &8.7 &0.85 \\
    \Lambda     &1.116 &0.0  &\N   &\N   & \N & \N  & \N &\N    &\N  & \Nc\\
    \Sigma      &1.189 &0.0  &\N   &\N   & \N & \N  & \N &\N    &\N  & \Nc\\
    \hline
    \res{S}{11}(1535) &1.525 &0.0  &0.6  &-0.60& \N & 0.5 & \N &0.1   &0.0 &2.2 \\
    \res{S}{11}(1650) &1.690 &0.030&1.0  &-0.45& \N &-0.45& \N &-0.1  &0.0 &-0.8\\
    \res{S}{31}(1620) &1.630 &0.100&3.7  &-0.12& \N &-0.12& \N & \N   &-0.8& \Nc\\
    \hline
    \res{P}{11}(1440) &1.520 &0.200&5.5  & 0.65& \N &0.65 & \N &0.0   &-2.0&0.0 \\
    \res{P}{11}(1710) &1.850 &0.300&3.0  & 0.25& \N &-0.8 & \N &0.0   &-3.0&2.0 \\
    \res{P}{13}(1720) &1.750 &0.300&0.12 &-0.75&0.25&-0.25&0.05&-0.035&0.0 &0.12\\
    \res{P}{33}(1230) &1.230 &0.0  &1.7  &-2.2 &-2.7&-2.2 &-2.7& \N   &0.0 & \Nc\\
    \res{P}{33}(1600) &1.855 &0.150&0.0  &-0.4 &-0.6&-0.4 &-0.6& \N   &0.55& \Nc\\
    \hline
    \res{D}{13}(1520) &1.515 &0.050&1.2  & 2.6 & 2.5& 2.6 & 2.5&2.0  &0.0 &1.2 \\
    \res{D}{13}(1700) &1.700 &0.090&0.0  &-0.5 & 0.0&-0.5 & 0.0&0.0  &0.3 &-0.04\\
    \res{D}{33}(1700) &1.670 &0.250&0.8& 1.5 & 0.6  & 1.5 & 0.6& \N  &-3.0& \Nc\\
  \end{tabular}
  \end{ruledtabular}
\end{table}
}

To be more specific, the loop corrections generated in the \K-matrix procedure
include only those diagrams which correspond to two on-mass-shell particles
in the loop~\cite{kon00,kon02}. This is the minimal set of diagrams one has to
include to ensure two-particle unitarity. Not included, thus, are all diagrams
that are not two particle reducible. This excludes the $\gamma N \to 2\pi N$
channels from the realm of our model. In addition, only the convergent pole
contributions i.e.\ the imaginary parts of the loop correction, are generated.
The omitted real parts are important to guarantee analyticity of the
amplitude and may have complicated cusp-like structures at energies where
other reaction channels open. In principle, these can be included as form
factors as is done in the dressed \K-matrix procedure~\cite{kon00,kor03}.
For reasons of simplicity we have chosen to work with purely phenomenological
form factors in the present calculations. An alternative procedure to account
for the real-loop corrections is offered by the approach of Ref.~\cite{sat96}
which is based on the use of a Bethe-Saltpeter equation. This approach was
recently extended to kaon production in Ref.~\cite{chi04}. Another possible
approach is the one discussed in Ref.~\cite{lut02} which is based on an N/D
expansion of the $T$ matrix combined with a dispersion-integral approach.

The strength of the \K-matrix procedure is that in spite of its simplicity,
several symmetries are obeyed by it~\cite{sch02}. As was already noted the
resulting amplitude is unitary provided that \K\ is Hermitian, and it obeys
gauge invariance provided the kernel is gauge-invariant. In addition, the
scattering amplitude complies with crossing symmetry when the kernel is
crossing symmetric. This property is crucial for a proper behavior in the
low-energy limit~\cite{kon02,kon04} of the scattering amplitude.
Coupled-channels effects are automatically accounted for by this approach for
the channels explicitly included into the \K-matrix as the final states.

As a result of this channel coupling, the resonances generate widths which
are compatible with their decays to channels included in the model space.
For some resonances, such as the $\Delta$ and the $S_{11}$(1535), this 
corresponds to their total width. Other resonances, particularly the high 
lying ones, may have important decay branches to states that are not included
in the model basis. To account for this in our calculations, we have added 
an explicit dissipative part to the corresponding propagators. The magnitudes
of these widths are equivalent to decay widths of the resonances to states 
outside of our model space.

The resonances which are taken into account in building the kernel are
summarized in \tblref{resonances}. In the current work we limit ourselves
to the spin-$\frac12$ and spin-$\frac32$ resonances as in this energy regime
higher spin resonances are known~\cite{shk07} to give only a minor contribution
to the $\eta N$ channel which is of primary interest here. Spin $\frac32$
resonances are included with so-called gauge-invariant vertices which have the
property that the coupling to the spin-$\frac12$ pieces in the Rarita-Schwinger
propagator vanish~\cite{pas00,kon00}. We have chosen this prescription since
it reduces the number of parameters as we do not have to deal with the
off-shell couplings. The effects of these off-shell couplings can be absorbed
in contact terms \cite{pas01} which we prefer, certainly within the context of
the present work.

The masses of the resonances given in \tblref{resonances} are bare masses and
they thus may deviate from the values given by the Particle Data Group
\cite{PDG}. Higher-order effects in the \K-matrix formalism do give
rise to a (small) shift of the pole-position with respect to the bare masses.
The masses of very broad resonances, in particular the $P_{11}$, are not well
determined - values lying in a broad range (typically a spread of the order of
a quarter of the width) give comparable results.  The width quoted in
\tblref{resonances} corresponds to the partial width for decay to states
outside our model space. The parameters as quoted in \tblref{resonances} are
mostly unchanged as compared to those presented in previous calculations within
this model~\cite{uso05,uso06}. It should however, be noted that values of the
coupling constants for the electromagnetic vertices in Ref.~\cite{uso06} were
given after multiplying them (mistakenly) by a factor of 2. This has not been done in
\tblref{resonances}. The $t$-channel contributions which are included in
the kernel, are summarized in \tblref{particles}.
\begin{table}
  \caption{ \tbllab{particles}
    Mass, spin, parity and isospin of the mesons which are
    included in the model. The rightmost column 
    specifies in which reaction channels their $t$-channel
    contribution are taken into account. }
  \begin{ruledtabular}
  \begin{tabular}{C|d|CC|c}
    \mc{Meson}&\mc{M [GeV]}& S^\pi & I & t-ch contributions \\
    \hline
    \pi     & 0.135     & 0^- & 1 & (\ch \gamma N \phi N), (\ch \pi N \rho N)\\
    K       & 0.494     & 0^- &\oh& (\ch \gamma N K\Lambda), (\ch \gamma N K\Sigma) \\
    \phi    & 1.019     & 1^- & 0 & \\
    \eta    & 0.547     & 0^- & 0 & (\ch \gamma N \phi N) \\
    \hline
    \rho    & 0.770     & 1^- & 1 & (\ch \gamma N \pi N), (\ch \gamma N \eta N),                                               (\ch K\Lambda K\Sigma), \\
            &           &     &   & (\ch K\Sigma K\Sigma), (\ch N\pi K\Lambda), \\
            &           &     &   & (\ch N\pi N\eta), (\ch N\pi N\pi) \\
    \omega  & 0.781     & 1^- & 0 & (\ch N\gamma N\pi),(\ch N\gamma N\eta) \\
    \sigma  & 0.760     & 0^+ & 0 & (\ch N\gamma N\phi), (\ch N\pi N\pi) \\
    K^*     & 0.892     & 1^- &\oh& (\ch N\gamma K\Lambda),(\ch N\gamma K\Sigma), \\
            &           &     &   & (\ch K\Lambda N\eta), (\ch K\Sigma N\eta), \\
            &           &     &   & (\ch N\pi K\Sigma) \\
  \end{tabular}
  \end{ruledtabular}
\end{table}

In the present calculation all primary coupling constants to the nucleon
have been chosen to be positive. In particular, the sign of $g_{NK \Lambda}$
deviates from the customary negative value~\cite{cot04} (see
\tblref{resonances}). In a calculation like ours and many of those cited in
Ref.~\cite{cot04} this sign is undetermined. Changing the sign of all coupling
constants involving a single $\Lambda$-field leaves the calculated observables
invariant since it corresponds to a sign redefinition of this field. In weak
decay the ratio of the vector to axial-vector coupling does correspond to an
observable. The magnitudes of the couplings are within the broad range
specified in~\cite{cot04}.

\subsection{Model space, channels included}

To keep the model manageable and relatively simple, we consider only stable
particles or narrow resonances in two-body final states which are important
for $\eta$-meson  photoproduction. The $\Lambda K$, $\Sigma K$, $N\phi$,
$N\eta$ and $N\gamma$ are the final states of primary interest, and the
$N\pi$ final state is included for its strong coupling to most of the
resonances. Three-body final states, such as $2\pi N$, are not included
explicitly for reasons of simplicity. Their influence on the width of
resonances is taken into account by assigning an additional (energy dependent)
width to them~\cite{kor98}. To investigate the effects of coupling to more
complicated states, we have also included the $N\rho$ final state. As was
shown in Ref.~\cite{uso05}, inclusion of the $\rho$ channel has a strong
influence on the pion sector but only a relatively minor effect on $\Lambda$
and $\Sigma$ photoproduction.

The components of the kernel which couple the different non-electromagnetic
channels are taken as the sum of tree-level diagrams, similar to what is used
for the photon channels. For these other channels no additional parameters
were introduced and they thus need no further discussion.

\section{Form-factors \& gauge restoration \seclab{ff}}

A calculation with Born contributions, without the introduction of form
factors, strongly overestimates the cross section at higher energies.
Inclusion of coupled-channels effects reduces the cross section at high
energies; however not sufficiently to produce agreement with the experimental
data, and one is forced to quench the Born contribution with form factors.
There are two physical motivations for introducing form factors (or vertex
functions) in our calculation.  First of all, at high photon energies one may
expect to become sensitive to the short-range quark structure of the nucleon.
Because this physics is not included explicitly in our model, we can only 
account for it through the introduction of phenomenological vertex functions.
The second reason has to do with the intermediate-range effects because of 
meson-loop corrections which are not generated through the \K-matrix 
formalism. Examples of these are given in Refs.~\cite{kon00,kor03}.

In our approach as well as that of Ref.~\cite{pen02}, the form-factors are
not known \emph{a priori} and thus they introduce certain arbitrariness in
the model. In the current paper we limit ourselves to dipole form-factors in
$s$-, $u$-, and $t$-channels because of their simplicity,
\begin{equation}\eqlab{ff-dipole}
  F_m(s)=\frac{\lambda^2}{\lambda^2+(s-m^2)^2} \;,
\end{equation}
where $m$ is the mass of the propagating particle and $\lambda$ is the
cut-off parameter. For ease of notation we introduce the subtracted form 
factors
\begin{equation}\eqlab{ff-twiddle}
  \tf_m(s)=\frac{1-F_m(s)}{s-m^2} \,,
\end{equation}
where $F_m(s)$ is normalized to unity on the mass-shell, $F_m(m^2)=1$, and
$\tf_m(m^2)$ is finite.

However, only in the kaon sector we use a different functional form for the
$u$-channel form-factors
\begin{equation}\eqlab{ff-u-channel}
  H_m(u)=\frac{u\lambda^2}{\big( \lambda^2+(u-m^2)^2 \big)m^2}.
\end{equation}
The argumentation for this different choice is presented in the discussion
of the $\Sigma$-photoproduction results in Ref.~\cite{uso05}. Often a different
functional form and cut-off values are introduced for $t$-channel form factors.
Although this can easily be motivated, it introduces additional model
dependence and increases the number of free parameters. To limit the overall
number of parameters we have taken the same cut-off value ($\lambda=1.2\;
\mbox{GeV}^2$, see \eqref{ff-dipole}) for all form-factors except for Born
contributions in kaon channels where we used $\lambda=1.0\;\mbox{GeV}^2$.

Inclusion of form-factors will in general break electromagnetic
gauge-invariance of the model. Therefore, a gauge-restoration procedure should
be applied. In Ref.~\cite{uso05}, the implications of various gauge-restoration
procedures was studied for the $\gamma p \to K \Sigma $ amplitude. It was
observed that the gauge-invariance restoration procedure is model
dependent which may give rise to strongly different Born contributions
to the amplitude. Therefore, the choice of a procedure to be adopted is
guided by its ability to describe the experimental data.  It was found that
the gauge-restoration procedure of Davidson and Workman~\cite{dav01} provided
the best description of the data on the $K \Sigma$ photoproduction. We
have used this procedure in the present work also.

We note that fitting the pion-scattering and pion-photoproduction amplitudes 
fixes masses as well as pion- and photon-coupling constants for most of
the resonances. This limits strongly the number of free parameters for the 
kaon-production channels.

\section{Results and Discussions}

Our main aim in this paper is to use the comprehensive data base of the
$\eta$-meson production to check various ingredients and input parameters
of our unitary coupled-channels field theoretic model of meson production in
photon induced reactions on nucleons. The requirement of a simultaneous fit to
data for a multitude of observables is expected to provide a strong constraint
on the values of the model parameters. It is also likely to highlight the role
of channel couplings in various regions of photon energies because several
calculations of $\eta$-meson photoproduction reactions have neglected these
effects~\cite{ben95,tia99,wen02,try04,ani05,nak08}.
\begin{figure*}
\begin{center}
\includegraphics[width=0.5 \textwidth]{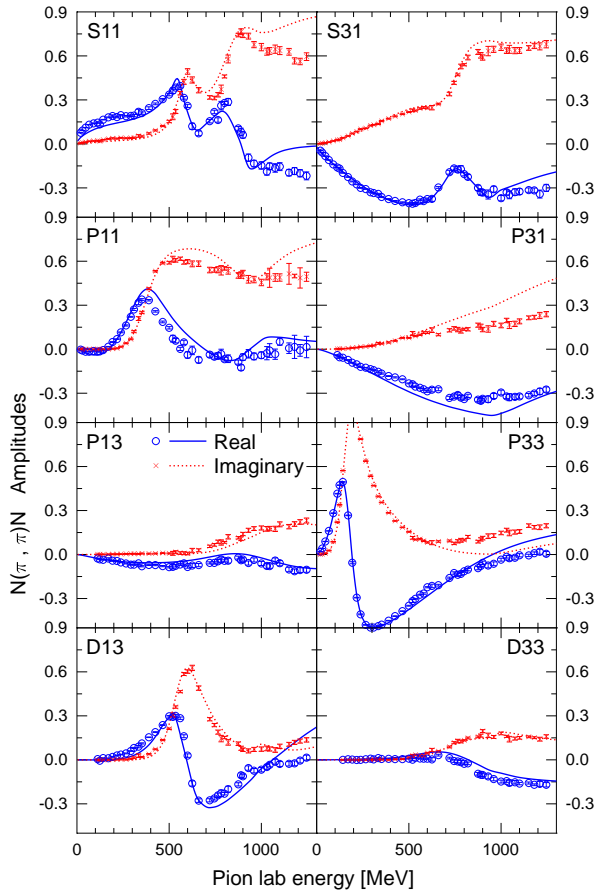}
\vskip -0.1in
\caption{[color online]
The real (solid lines) and imaginary (dotted lines) pion-nucleon $S$-, $P$-,
and $D$-wave amplitudes. The curves represent the results of our calculations
while open circles and crosses are the results of Virginia Tech single-energy 
partial wave FA08 analysis~\protect\cite{vir96}. 
}
\end{center}
\end{figure*}

We emphasize however, that even the large experimental data base may not allow
to fix the extracted parameters uniquely within the unitary coupled-channels
effective-Lagrangian model \cite{pen02}. This is due to the fact that it is
necessary to include empirical form factors in the model to regularize the
amplitudes at higher energies. These form factors require a gauge-invariance
restoration procedure which involves ambiguities. Nevertheless, confronting the
model with a large data base is expected to provide a means to overcome this
problem.

The parameters in the model have been adjusted~\cite{uso06} to reproduce the 
Virginia Tech partial wave amplitudes of Arndt et al.~\cite{vir96}. In Fig.~2
we present a comparison of our calculated pion-nucleon $S$-, $P$-, and 
$D$-wave amplitudes for isospins $I$ = 1/2 and 3/2 channels with those of
the FA08 single-energy partial wave amplitudes of Ref~\cite{vir96}. The 
corresponding results for pion photoproduction and the Compton scattering
are given in Ref.~\cite{uso06}. We see that both real and imaginary parts
of the pion-nucleon scattering amplitudes are described well although some
differences start to show up at the upper limit of the energy range
considered.

The data for $\eta$ meson photoproduction consist of total and differential 
cross sections measured at CB-ELSA at Bonn~\cite{cre05,bar07} on the proton, 
for photon energies ranging from 0.750~GeV to 3~GeV. These data, therefore, 
cover not only the entire resonance region but also the region where the 
background contributions ($t$-channel amplitudes mainly) are expected to be 
dominant. These data are consistent with the measurements reported by 
Mainz-TAPS~\cite{kru95}, CLAS~\cite{dug02} and GRAAL \cite{ren02,bar07a} 
collaborations where photon energies go up to 0.790 GeV, 1.95~GeV and 1.1~GeV,
respectively. The data on beam asymmetry have been taken by the CB-ELSA group 
for photon energies in the range of 0.800~GeV to 1.4~GeV.  In this case too 
there is an agreement (except for a single bin in photon energy) between these
data and those of the GRAAL group~\cite{aja98,bar07a} where photon energies 
are in the range to 0.724~GeV to 1.472~GeV. One set of data on the target 
asymmetry (analyzing power) has already been reported in Ref.~\cite{boc98} 
and more measurements are planned by the CB-ELSA group.
\begin{figure*}
\begin{center}
\includegraphics[width=0.5 \textwidth]{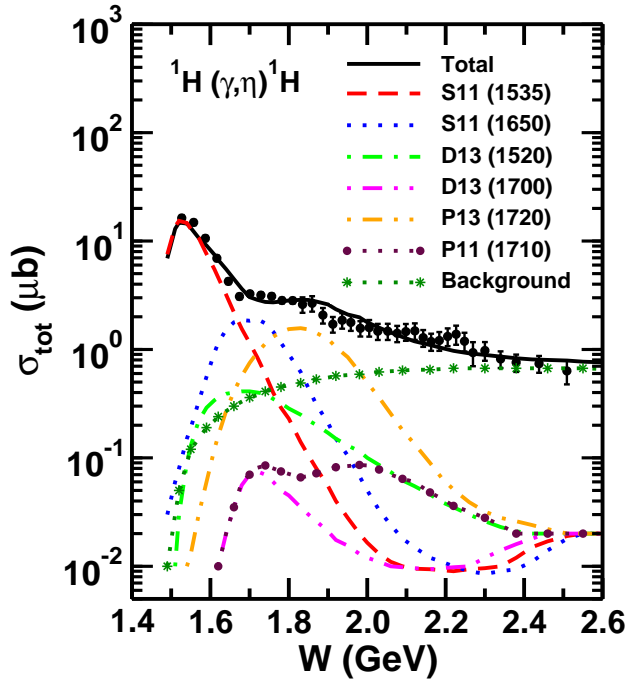}
\vskip -0.1in
\caption{[color online]
Total cross section for the $\gamma p \to \eta p$ process as a function
of $\gamma p$ invariant mass. The experimental data are taken from
Ref.~\protect\cite{cre05}. Contributions of resonances $S_{11}$(1535),
$S_{11}$(1650), $P_{11}$(1710),$P_{13}$(1720), $D_{13}(1520)$, $D_{13}$(1700)
are shown by various curves as indicated in the figure. Also shown are
the background contributions which consist of Born  and $u$- and $t$-channel
terms.
}
\end{center}
\end{figure*}

Data on $\eta$-meson photoproduction off the neutron are not as well developed
and extensive as those on the proton due to the non-availability of free
neutrons as targets. In most cases the photoproduction is measured on the
neutron bound in the deuteron by performing experiments on the deuterium
target. In this sense the corresponding cross sections are "quasi-free".
Calculations done for the free neutron will have to be corrected for the Fermi
motion and other nuclear effects before comparing them with the quasi-free
production data. This procedure is not unique because of the prescription
used in unfolding the recoil momentum of the "spectator" particle. Furthermore,
quasi-free cross sections are also influenced by off-shell and final-state
interaction effects. Nevertheless, quality data are now available
for total and differential cross sections for $\eta$-meson photoproduction
off the quasi-free neutrons~\cite{jae08}. This also includes the ratio of the
total cross sections for this reaction on the neutron and the proton measured
under identical conditions off the nucleon bound in the deuteron.

In Fig.~3, we show the contribution of various resonances to the total
cross section for the $\gamma p \to \eta p$ reaction which is plotted here as
a function of the $\gamma p$ invariant mass $W$. The experimental data are
taken from Ref.~\cite{cre05}. Since data of the various experimental
collaborations are consistent with each other, we show the data of only one
group (which span the maximum range of photon energy, E$_\gamma$) on this
plot. From this figure it is apparent that the general features of the data
are described reasonably well by our calculations in the entire 
range of beam energies where data are available. It should, however, be 
mentioned that our model slightly overpredicts (underpredicts) the data 
for $W$ in the range of 1.85-1.95 GeV (2.18-2.25 GeV). Even the 
partial wave analysis fits to these data reported in Ref~\cite{cre05} show 
signs of underpredicting the cross section for $W \sim$ 1.85-1.95. 
\begin{figure*}
\begin{center}
\includegraphics[width=0.8 \textwidth]{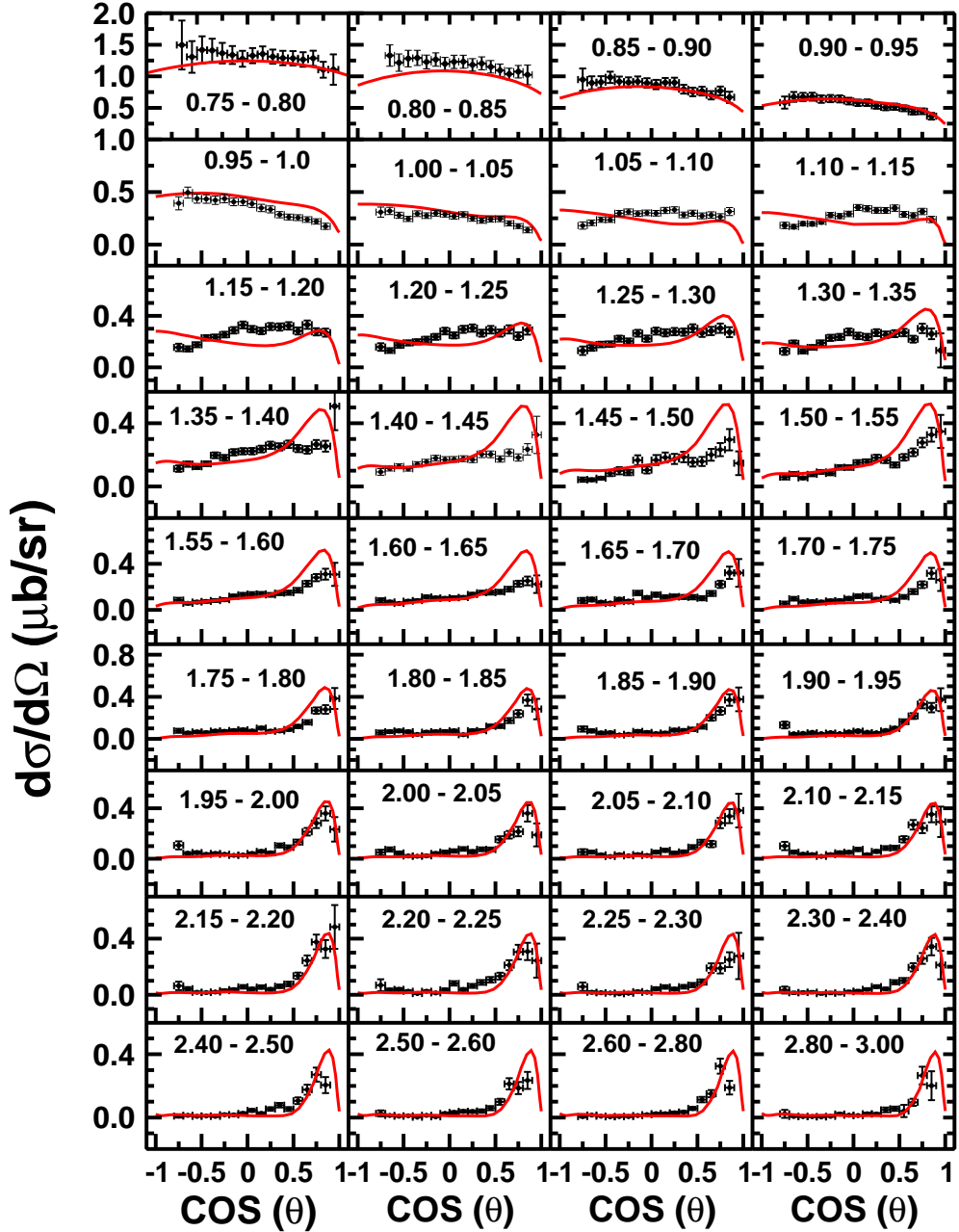}
\vskip -0.1in
\caption{[color online]
Differential cross section for the $\gamma p \to \eta p$ process as a function
of the cosine of the $\eta$ c.m.\ angle for 36 photon energy bins. The energy
bin is indicated in each graph in GeV. The experimental data are taken from
Ref.~\protect\cite{cre05}.
}
\end{center}
\end{figure*}

We note that while contributions of the $S_{11}$(1535) resonance dominate the
cross sections from near threshold to $W$ values of 1.7~GeV (corresponding to
E$_\gamma \sim$ 1.1~GeV), those of the $S_{11}$(1650) and $P_{13}$(1720)
resonance are important for E$_\gamma$ between 0.950~GeV to 1.28~GeV and 1.1~GeV
to 2.2~GeV, respectively. In fact, omission of the $S_{11}$(1650) resonance
worsens the description of the data for E$_\gamma$ in the range of 1.0-1.2~GeV.
The non-negligible contribution of this resonance is consistent with the
conclusions of Refs.~\cite{wen02,shk07,nak08}. However, this is in contrast to
the partial-wave analysis results of Ref.~\cite{ani05}. We further note that 
magnitudes of the $D_{13}$(1520), $D_{13}$(1700) and $P_{11}$(1710) resonances
are comparatively small in the entire range of photon energies.

It should be remarked that at $W\sim$ 1.716 GeV, there are some differences 
between the GRAAL~\cite{ren02} and ELSA~\cite{cre05} data - there is a bump 
in the former around this energy which is almost absent in the latter. The 
existence of a third $S_{11}$ resonance with a mass of 1.712~GeV was suggested
in Ref.~\cite{sag01} from the analysis of the preliminary GRAAL data. Very 
recently, calculations performed within a constituent quark model show that 
the inclusion of the third $S_{11}$ resonance with a slightly higher mass of
1.730 GeV, improves the agreement with the differential cross section and 
beam asymmetry data at some angles~\cite{jun08}. We, however, have chosen not
to include a third $S_{11}$ resonance in this work which is in line with the 
calculations done in Refs~\cite{wen02,shk07}. Anyhow, in principle
it is straight forward to include this resonance into our analysis which
we propose to do in a future study. 
\begin{figure*}
\begin{center}
\includegraphics[width=0.5 \textwidth]{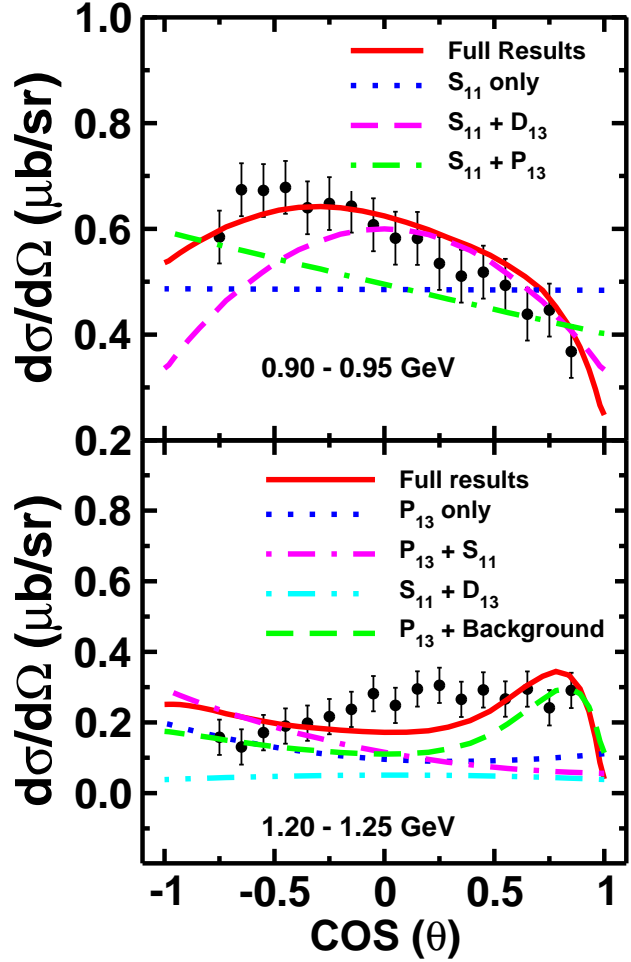}
\vskip -0.1in
\caption{[color online]
Contributions of various resonance terms to the differential cross sections
of the $\gamma p \to \eta p$ process as a function of the cosine of the 
$\eta$ c.m.\ angle for photon energy bins of 0.90-0.95~GeV (upper panel) 
and 1.20-1.25~GeV (lower panel). Various curves are explained in the figure. 
The experimental data are from Ref.~\protect\cite{cre05}.
}
\end{center}
\end{figure*}

Total cross sections beyond 2~GeV are almost solely governed by the 
contributions of the background terms which are dominated by the $t$-channel 
diagrams. In this region all resonance contributions are small and comparable
to each other. The vector meson ($\rho$ and $\omega$) exchange terms give the
largest contribution to the $t$-channel amplitudes. The importance of this 
mechanism in the pion- and photon-induced $\eta N$ production, was already 
emphasized in the first coupled-channels model~\cite{sau95} for these 
reactions.  There is some indication of a small bump like structure in the 
data for $W$ around 2.2~GeV which might indicate the presence of a resonance 
in this region. The partial-wave analysis of these data~\cite{ani05} does 
include a $P_{13}$ resonance with mass 2.2~GeV and total width of 0.360~GeV in
the fitting procedure. However, the existence of such a resonance is not yet 
to find a wider support (see, {\it e.g.}, Ref.~\cite{PDG}).

Differential cross sections (DCS) provide more valuable information about
the reaction mechanism~\cite{tia94}. They reflect the quantum number of the
excited state (baryonic resonance) when the cross section is dominated by it.
DCS include terms that weigh the interference terms of various components of
the amplitude with the outgoing $\eta$ angles. Therefore, the structure of
interference terms could highlight the contributions of different resonances
in different angular regions. For the $\gamma p \to \eta p$ reaction,
DCS data exist for 36 photon energy bins in the range of 0.750~GeV to 3.0~GeV
covering a wide range of $\eta$ center of mass (c.m.) angles ~\cite{cre05}.
In Fig.~4, we compare the results of our calculations for the angular
distributions with these data. We see that while our model describes the 
general trends of the data well in the complete energy region, a few specific 
details of the data are missed for some energy bins  in the region of 
1.0-1.60~GeV. For example, the curvature of the data is not reproduced by 
our calculations at middle angles for the photon energy bins lying between 
1.0-1.25 GeV. There is a overprediction of the data at very forward angles
for energy bins between 1.40-1.60 MeV. 
\begin{figure*}
\begin{center}
\includegraphics[width=0.5 \textwidth]{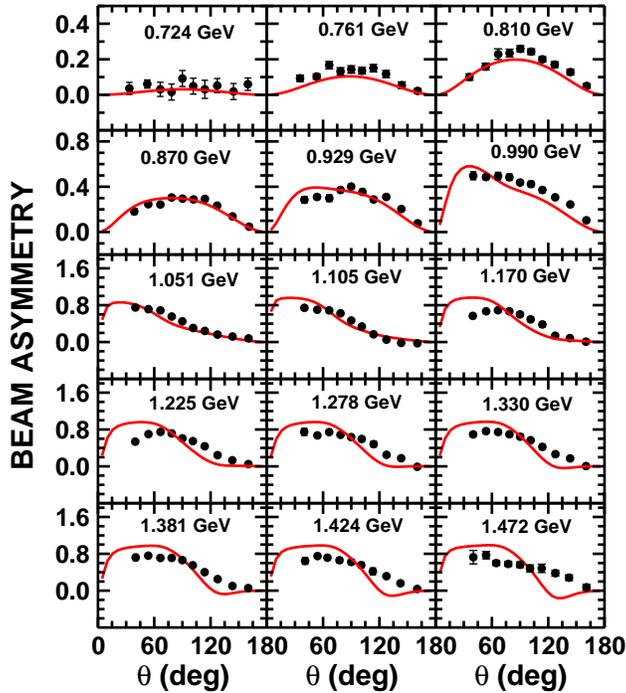}
\vskip -0.1in
\caption{[color online]
Beam asymmetry for the $\gamma p \to \eta p$ reaction as a function
of $\eta$ c.m.\ angle for 15 photon energies. The experimental
data are taken from Ref.~\protect\cite{bar07a}.
}
\end{center}
\end{figure*}

Although individual contributions of the $D_{13}$ and $P_{13}$ resonances to
the total cross section are rather small for E$_\gamma \leq$ 1.0~GeV, their
interference with the dominant $S_{11}$ resonance amplitudes are vital for
describing the experimental DCS data in this energy regime. We demonstrate
this in Fig.~5 where we show individual contributions of various resonance
terms to differential cross sections for photon energy bins of 0.900-0.950~GeV
(upper panel) and 1.200-1.225~GeV (lower panel). For the lower energy bin,
as expected, the contributions of $S_{11}$ resonances are almost flat. However,
inclusion of the $D_{13}$ and $P_{13}$ resonances together with $S_{11}$ is
crucial for describing the data for $\eta$ angles below and above $90^\circ$,
respectively. For the higher energy bin (lower panel), the $S_{11}$ and
$D_{13}$ resonances do not contribute much at forward angles - here the
$P_{13}$ resonance and the background terms put together reproduce the shapes
and magnitudes of the measured DCS. The forward peaking of the angular
distributions beyond 1.8~GeV reflects the dominance of the $t$-channel meson
exchange diagrams.
\begin{figure*}
\begin{center}
\includegraphics[width=0.3 \textwidth]{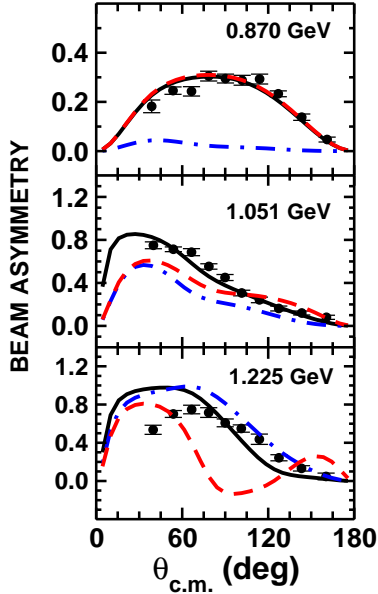}
\vskip -0.1in
\caption{[color online]
Beam asymmetry for the $\gamma p \to \eta p$ reaction as a function
of $\eta$ c.m.\ angle for photon energies of 0.870~GeV, 1.051~GeV and
1.225~GeV. Solid lines show the full calculations where all the considered
resonances and the background terms are included. The dashed and dashed-dotted
curves are obtained when the $P_{13}$(1720) and the $D_{13}$(1520) resonances are
omitted, respectively. The experimental data are taken from the
Ref.~\protect\cite{bar07a}.
}
\end{center}
\end{figure*}

Polarization observables are more sensitive to the contributions of the
resonances which are not dominant in the cross sections. Beam asymmetry
($\Sigma_B$) is the measure of the azimuthal anisotropy of a reaction yield
relative to the linear polarization of the incoming photon. As compared to
the cross-section data this observable is less sensitive to the $S_{11}$(1535)
resonance even at energies close to threshold. In Fig.~6, we compare
results of our calculations for $\Sigma_B$ with the experimental data of Ref.
\cite{bar07a} which is available for 15 values of $E_\gamma$ in the energy
range of 0.724~GeV to 1.5~GeV. We note that there is an overall agreement
between our calculations and the experimental data in the complete range of
photon energies.

The sensitivity of individual resonances to $\Sigma_B$ is studied in Fig.~7
for three representative photon energies of 0.870~GeV, 1.051~GeV and 1.225~GeV. In each case the solid, dashed and dashed-dotted lines represent results of
full calculations, and those obtained by ignoring contributions of the
$P_{13}$(1720) and $D_{13}$(1520) resonances, respectively. We note that at the
near threshold photon energy of 0.870~GeV, the $D_{13}$(1520) resonance plays a
very crucial role - the beam asymmetry goes from its maximum value of 0.35 to
almost zero if this resonance is ignored. At the larger photon energy of 1.225
GeV, contributions of the $P_{13}$(1720) resonance are important since the data
can not be described without including them. At the intermediate photon energy
(1.051~GeV) both the $D_{13}$(1520) and the $P_{13}$(1720) resonances
\begin{figure*}
\begin{center}
\includegraphics[width=0.5 \textwidth]{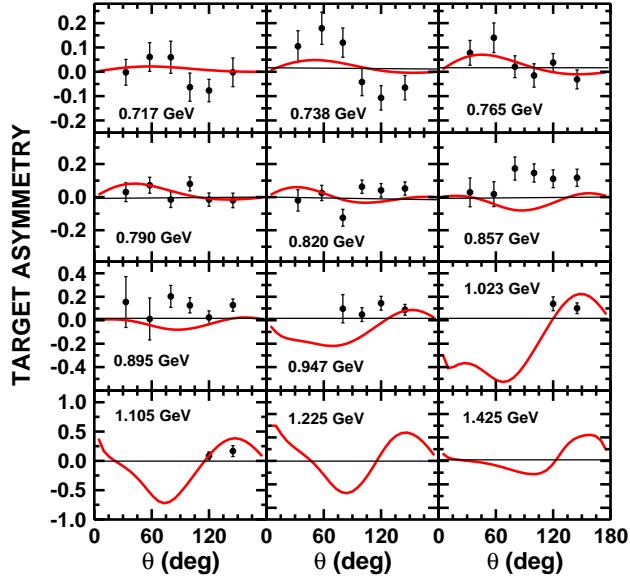}
\vskip -0.1in
\caption{[color online]
Target asymmetry for the $\gamma p \to \eta p$ reaction as a function of $\eta$
c.m.\ angle for various photon energies. Solid lines are the full calculations
where all the considered resonances and the background terms are included.
Experimental data are taken from the Ref.~\protect\cite{boc98}.
}
\end{center}
\end{figure*}
\noindent
are important. Our results for the higher photon energy (1.225~GeV) are in
agreement with those of the Bonn-Gatchina partial wave analysis \cite{ani05} as
reported in Ref.~\cite{els07} for the photon energy bin of (1.250 $\pm$ 
0.050)~GeV (in this reference the individual contributions of resonances are 
shown only for this bin). In contrast to this, in the eta-MAID 
analysis~\cite{els07}, $\Sigma_B$ is insensitive to the $P_{13}$ resonance at
this value of $E_\gamma$. The crucial role of the $D_{13}$(1520) resonance in
describing the beam asymmetry at lower photon energies, was demonstrated 
already in Ref.~\cite{tia94}. Our work establishes this in conjugation with 
the description of the experimental data for the first time.

In Fig.~8, we have compared the results of our calculations for the target
asymmetry ($TA$) with the corresponding experimental data reported in Ref.
~\cite{boc98}. In general the predictions of our model are consistent with
the trends seen in the data within error bars, except for $E_\gamma$ between
0.857~GeV and 0.947~GeV where there is some incompatibility between data and
our results. While at 0.857~GeV and 0.895~GeV the sign of the data is missed,
at 0.947~GeV the sign change in the theoretical results occurs at somewhat 
higher angles in comparison to that of the data. In any case, the comparison 
between theory and data should be viewed in the light of the fact that the 
data have rather large error bars in both $\eta$ angles and $E_\gamma$. It is
worthwhile to note that our results show a qualitative similarity with those
of Ref.~\cite{pen02} even though at certain photon energies energies 
some differences are noticeable between the two calculations. 
\begin{figure*}
\begin{center}
\includegraphics[width=0.5 \textwidth]{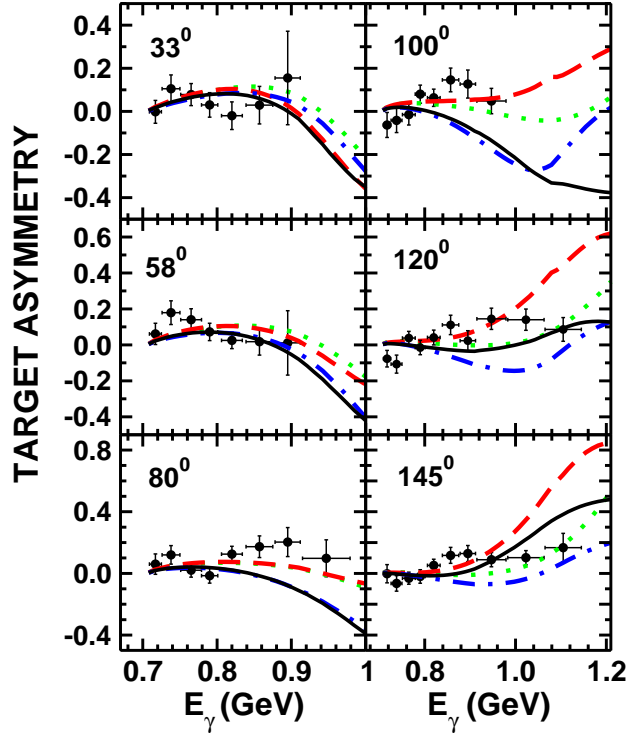}
\vskip -0.1in
\caption{[color online]
Target asymmetry for the $\gamma p \to \eta p$ reaction as a function of photon
energy for $\eta$ c.m.\ angles of 33$^\circ$, 58$^\circ$, 80$^\circ$,
100$^\circ$, 120$^\circ$, 145$^\circ$. Solid lines show full results (as in
Fig.~8) while dotted and dashed-dotted curves are obtained when $P_{13}$ and
$D_{13}$ resonances are omitted from the calculations, respectively. The
dashed curves result when both the $P_{13}$ and the $D_{13}$ resonances are 
excluded.  The experimental data are taken from~\protect\cite{boc98}.
}
\end{center}
\end{figure*}

In Fig.~9 we show $TA$ as a function of $E_\gamma$ for various $\eta$ angles. In
this figure we also show results obtained by omitting the $P_{13}$ and $D_{13}$
resonance contributions from the calculations (shown by dotted and
dashed-dotted lines, respectively). Dashed curves show results where both
these resonances have not been included. At lower energies for all angles, the
interference effects of the $P$ and $D$ resonances with the dominant $S_{11}$
ones are small. However, as energy increases, full results (solid
lines) start deviating from those obtained by retaining only the $S_{11}$
resonances, particularly for angles $> 33^\circ$. It is to be noted that
interference effects of both the $P_{13}$ and the $D_{13}$ resonances with the
$S_{11}$ are important at higher energies and angles. In contrast to our study,
the interference terms of the $P$ resonances were not included in the analysis
of Ref.~\cite{tia99}. In general our full results reproduce the trends seen 
in the data for all angles except for 80$^\circ$ and 100$^\circ$ where there 
are discrepancies between the two at higher photon energies.

The effects of channel couplings on the total cross section and beam asymmetry
are studied in Figs.~10 and 11, respectively. In Fig.~10, we compare the
results of full coupled channel calculations for the total cross section of the
$\gamma p \to \eta p$ reaction (solid line) with those obtained by switching
off the channel coupling effects (this will be referred as NCC) (dashed line).
Full calculations are the same as those shown in Fig.~3. In the NCC case, the
amplitudes of various processes are simply added together, ignoring the
modifications to the widths of the resonances introduced by the channel
couplings. We notice that for W $>1.8$~GeV the differences between the full and
the NCC results are very small. However, at lower energies, the channel-coupling
effects are large and are crucial for describing the data. In fact, at some
energies, the resonance propagators can develop poles in the absence of
channel couplings. Thus modifications introduced to the widths of the
resonances due to channel couplings are indeed vital for reproducing the
energy dependence of the experimental cross sections.
\begin{figure*}
\begin{center}
\includegraphics[width=0.3 \textwidth]{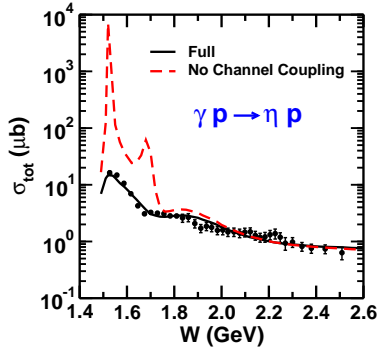}
\vskip -0.1in
\caption{[color online]
Effect of channel coupling on the total cross section for the
$\gamma p \to \eta p$ reaction as a function of $\gamma p$ invariant mass.
The experimental data are taken from Ref.~\protect\cite{cre05}. The solid
line represents the results of the full coupled-channel calculations (same as
that in Fig.~3) while the dashed line shows the one where channel coupling
is switched off.
}
\end{center}
\end{figure*}

In Fig.~11, we show the effect of channel coupling on the beam asymmetry
$\Sigma_B$ as a function of the $\eta$ c.m. angle for various photon energies.
In this case too, we notice that channel coupling effects are vital for
describing the data at lower photon energies. For $E_\gamma < $ 1.225~GeV, the
$\Sigma_B$ in the NCC case are generally smaller and may even have wrong signs
as compared to those obtained in the full model. Thus channel coupling effects
are extremely important in describing the magnitudes and the relative signs of
the beam asymmetry data. The reason for this lies in the fact that polarization
observables are generally very sensitive to the imaginary parts of the
amplitudes which are governed by coupling to other channels via the optical
theorem.

For the excitation of an isospin-$\frac{1}{2}$ resonance, both isoscalar and
isovector components of the photon can contribute. In order to get information
about the amplitudes corresponding to these components, data are needed for
$\eta$-meson photoproduction off the neutron together with that off the proton.
The non-existence of the free neutron target has prompted the use of a
deuteron target to get information about the $\gamma n \to \eta n$ reaction.
However, experiments on the deuteron provide information about the quasi-free
production as the reaction takes place on the neutron bound in the deuteron
where the Fermi motion of the nucleon inside the deuteron strongly influences
the kinematics of the process.
\begin{figure*}
\begin{center}
\includegraphics[width=0.5 \textwidth]{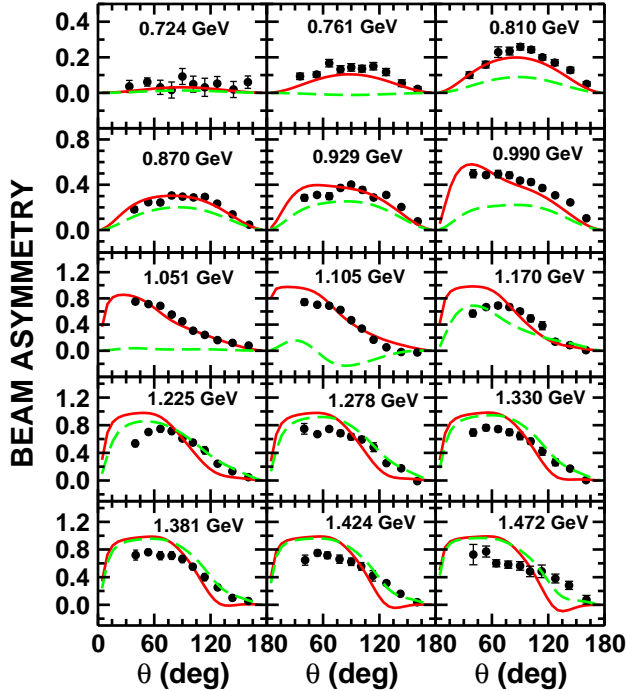}
\vskip -0.1in
\caption{[color online]
Effect of channel coupling on the beam asymmetry for the 
$\gamma p \to \eta p$ reaction as a function of $\eta $ c.m.\ angle for 
photon energies as indicated in each graph. Full coupled channel (same as 
that in Fig.~6)  and no channel couplings results are shown by solid and 
dashed lines, respectively. The experimental data are taken from 
Ref.~\protect\cite{bar07a}.
}
\end{center}
\end{figure*}

The TAPS collaboration has studied the quasi-free $\eta$ production off the
neutron for $E_\gamma$ ranging from threshold upto 0.820~GeV~\cite{wei03}. They
have reported a constant ratio of 2/3 for $\gamma n \to \eta n$ and
$\gamma p \to \eta p$ reactions at these near threshold energies.  At the GRAAL
facility both quasi-free production reactions have been explored 
simultaneously in the same experimental run with $E_\gamma$ going upto
1.6~GeV~\cite{kuz07}. They have reported a larger value for this ratio. 
Very recently, at the Bonn ELSA facility, simultaneous measurements have been
performed for these reactions for incident photon energies being as large as
2.5~GeV~\cite{jae08}. In this experiment data have been taken for angular 
distributions and total cross sections of the two reactions. The excitation 
function for $\eta$-meson photoproduction off the (quasi-free) neutron shows
a pronounced bump-like structure at the invariant mass of 1.68~GeV 
(corresponding to $E_\gamma \approx$ 1.1~GeV). A similar structure
was also seen in the GRAAL data~\cite{kuz07}. On the other hand, such a
structure is not seen in the reaction off the proton (see Fig.~3).

As described earlier, this  structure has been interpreted in altogether
different ways by different authors. We believe that before going for more
exotic explanations, conventional mechanisms for $\eta$ photoproduction on
both the neutron and the proton should first be investigated in detail. Since
our model provides a reasonable description of both cross section and
polarization data for the $\gamma p \to \eta p$ reaction, it is natural 
to use it to describe the $\gamma n \to \eta n$ data as well. This will help 
in determining the neutron helicity amplitudes of the relevant resonances.
\begin{figure*}
\begin{center}
\includegraphics[width=0.3 \textwidth]{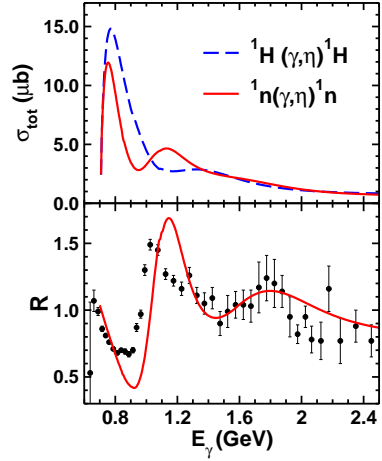}
\vskip -0.1in
\caption{[color online]
(Upper Panel) Total cross sections ($\sigma_{tot}$) for $\gamma p \to \eta p$
(dashed line) and  $\gamma n \to \eta n$ (full line) reactions as a function of
incident photon energy. The former is the same as that shown by full line in
Fig.~3. (Lower Panel) The ratio $R$ of total cross sections of the
$\gamma n \to \eta n$ and $\gamma p \to \eta p$ reactions as a function of 
photon energy. The data points shown in the lower panel have been taken from
the Ref.~\protect\cite{jae08}.
}
\end{center}
\end{figure*}

In Fig.~12, we present our results for the total cross sections of the
$\gamma p \to \eta p$ and $\gamma n \to \eta n$ reactions (upper panel) and for
their ratio ($R$) (lower panel) using the helicity couplings for resonances as
shown in Table I. We note that the value of the ratio $g_{n\gamma}/g_{p\gamma}$
is -0.8 for the $S_{11}$(1535) resonance which is in agreement with the 
results of the combined theoretical studies of the $\gamma p \to \eta p$ and 
the $\gamma n \to \eta n$ reactions~\cite{sau95,dre99,muk95,wen02,shk07} and 
also with those extracted from the experimental data on the ratio of the two 
cross sections~\cite{wei03,jae08}. In our study the helicity couplings of the 
neutron and the proton on the $S_{11}$(1650) and $D_{13}$(1520) resonances are
the same. However, for the $P_{13}$(1720) and the $P_{11}$(1710) resonances 
they differ from each other which was also the case in Ref.~\cite{shk07}.

We have not put the experimental data in the upper panel of Fig.~12 because
for a comparison between theory and the data on the quasi-free $\eta$
production, Fermi folding of our calculated cross sections will have to be
performed in a way similar to that implemented in the extraction of the
experimental data.  However, in the lower panel we do show the data which
are taken from Ref.~\cite{jae08}. We see that our calculations are consistent
with the experimental observation of a bump-like structure around a photon
energy of 1.1~GeV. Furthermore, yet another (rather broad) bump like structure
is seen in the measured $R$ at the photon energy of about 1.8~GeV. Our
calculations are compatible with this structure as well. Thus, interference
effects of mostly $S_{11}$(1535), $S_{11}$(1650), $P_{11}$(1710), and
$P_{13}$(1720) resonances within a coupled channel approach can lead to the
bump-like structures in the excitation function of $\eta$ photoproduction
off the neutron around photon energies of 1.1~GeV. They also lead to such
structures in the neutron to proton cross-section ratio. The structures seen
in our calculations are somewhat more pronounced than those observed in
Ref.~\cite{shk07} where they result from similar interference effects.
\begin{figure*}
\begin{center}
\includegraphics[width=0.5 \textwidth]{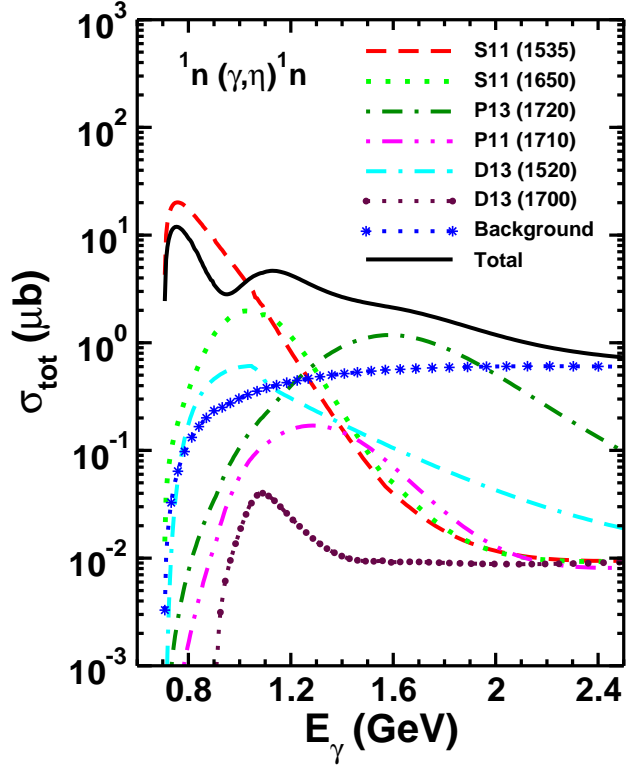}
\vskip -0.1in
\caption{[color online]
Total cross section for the $\gamma n \to \eta n$ reaction as a function
of incident photon energy. Contributions of the $S_{11}$(1535),
$S_{11}$(1650), $P_{11}$(1710),$P_{13}$(1720), $D_{13}(1520)$, $D_{13}$(1700)
resonances are shown by various curves as indicated in the figure. 
Also shown are the background contributions. 
}
\end{center}
\end{figure*}

The magnitude of the second peak seen in the total cross section of the
$\gamma n \to \eta n$ reaction, is very sensitive to the values of the neutron
helicity amplitudes of the $S_{11}$, $P_{11}$(1710) and $P_{13}$(1720)
resonances. In Fig.~13, we show the contributions of various resonant
states to the total cross section of this reaction. We note that in the
region of the second peak, the main contributing resonances are $S_{11}$(1535),
$S_{11}$(1650), $P_{13}$(1720), $P_{11}$(1710) and $D_{13}$(1520). Therefore,
the comparison of the  calculations (with Fermi folding) with the data on the
total production cross section of this reaction will provide a check of the
corresponding helicity amplitudes. Another interesting aspect of this figure
is that there is a strong negative interference among the resonances at lower
photon energies.  In this region the $S_{11}$(1535), $S_{11}$(1650), and
$D_{13}$(1520) resonances are most relevant. Therefore, the data can be used
to get further constraints on the helicity amplitudes of these three resonances.

As noted in case of the $\gamma p \to \eta p$ reaction, the differential cross
sections provide a more stringent constraint on the contributions of even those
resonances which participate only weakly in the total cross section. Data are
also reported in Ref.~\cite{jae08} on differential cross sections of the
$\gamma n \to \eta n$ reaction. In Fig.~14, we show results of our calculations
for differential cross sections of this reaction at several photon energies.
We have not put the data points in this figure as a meaningful comparison
between calculations and the data would require the Fermi folding of the
theoretical results as is stated before. However, some features
of our results are worth noticing. For $E_\gamma \leq 0.900$ GeV there is a
shape inversion in the angular distribution as compared to that observed in
the case of the $\gamma p \to \eta p$ reaction at similar values of $E_\gamma$.
This is consistent with the trend seen in the data. For $E_\gamma =$ 1.0~GeV
also similar inversion effect is present in the data except for the peaking 
in the forward direction.  Our calculations also have this feature. For photon
energies higher than this, angular distributions are of similar shapes for
the two reactions due to the dominance of the $t$-channel contributions in
both the cases.
\begin{figure*}
\begin{center}
\includegraphics[width=0.5 \textwidth]{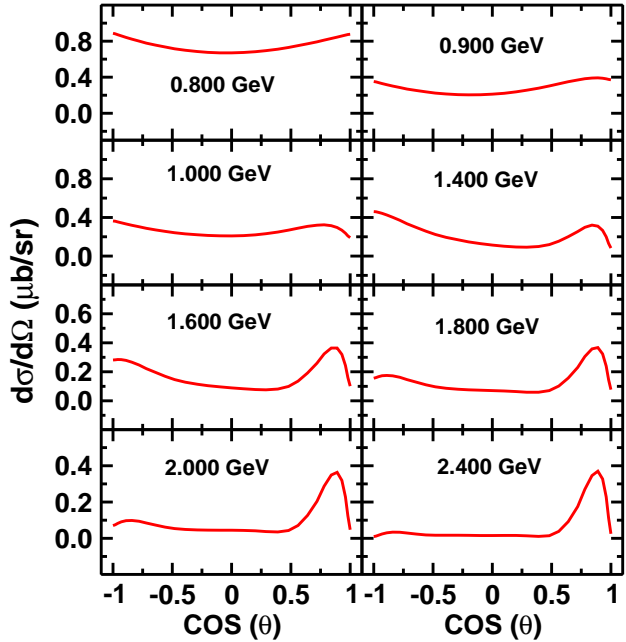}
\vskip -0.1in
\caption{[color online]
Differential cross section for the $\gamma n \to \eta n$ reaction as a function
of the cosine of the $\eta$ c.m. angle at several photon energies indicated
in each box.
}
\end{center}
\end{figure*}

\section{Summary and Conclusions}

In this paper we investigated the photoproduction of $\eta$ meson off nucleons
within a coupled-channels effective-Lagrangian approach which is based on the
\K-matrix method. Unitarity effects are correctly taken into account, since all
important final channels (consisting of two-body systems $\pi N$, $\eta N$,
$\phi N$, $\rho N$, $\gamma N$, $K \Lambda$, and $K \Sigma$) are included in
the \K-matrix kernel. We build this kernel by using effective Lagrangians for
the Born, $u$-channel, $t$-channel, and spin-$\frac{1}{2}$ and
spin-$\frac{3}{2}$ resonance contributions. Thus, the background contributions
are generated consistently and crossing symmetry is obeyed. The advantage
of a full coupled-channel calculation is that it allows for the simultaneous
calculation of observables for a large multitude of reactions with considerably
fewer  parameters than would be necessary if each reaction channel were
fitted separately. More importantly, the implementation of unitarity ensures
that the imaginary parts of the amplitudes are compatible with the cross
sections for other channels.

We showed that it is essential to use a full coupled-channels approach to
describe the meson-production reactions. The effects of channels coupling
are not merely a smooth change of the energy dependence of the cross sections.
In their absence, they can acquire structures that might be
misinterpreted as resonances. The polarization observables are also strongly
affected by these effects where omitting channels coupling leads to wrong
signs.

For $\eta$ photoproduction off the proton, our model provides a reasonable
description of the experimental data on total and differential cross
sections as well beam and target asymmetries for photon energies ranging from
threshold to up to 3~GeV. The previous effective Lagrangian based
coupled-channels calculations of this reaction were restricted to photon
energies below 2~GeV. We showed that the data on differential cross sections,
and beam and target asymmetries are very sensitive to contributions of
resonances which contribute only weakly to the total cross sections. For
these observables, the interference of these resonances with the dominant
$S_{11}$(1535) amplitudes is vital for describing the data even at lower
photon energies.

The second peak seen in the excitation function of the total cross section
for $\eta$ photoproduction off the neutron at photon energies around 1.1~GeV
can be explained by the interference effects of the $S_{11}$(1535),
$S_{11}$(1650), $P_{11}$(1710) and $P_{13}$(1720) resonances - there is no need
to introduce an exotic narrow resonance state. We find that the ratio of the
neutron to proton helicity amplitudes for the $S_{11}$(1535) resonance has to
be $\approx$ -0.8 in order to get the second peak in the total cross section
as seen in the data. Our calculations are also compatible with the broad 
bump-like structures observed at photon energies around 1.1~GeV and 1.8~GeV
in the ratio of total cross sections of $\eta$ photoproduction off the
neutron and the proton.

\section{Acknowledgments}
One of us (R.S.) acknowledges support from a visitors grant from the Dutch
Organization for Scientific Research (NWO) during his stay at KVI.

\appendix
\section{Effective Lagrangians}

We list here the effective Lagrangians for various vertices. $p$, $k$,
$p\prime$ and $-q$ represent four momenta of the initial nucleon, final meson,
final nucleon and photon, respectively. We assume that meson momenta are
directed into the vertex, so that energy momentum conservation reads as
$p + k = p\prime - q $.

For the nucleon vertices the following couplings were used
\begin{equation}\eqlab{LagrangianN}
  \begin{aligned}
    \La_{NN\pi} &= i g_{NN\pi}{\bar{\Psi}}_N \YYv{\vecvarphi_\pi}\Psi_N \\
    \La_{NN\eta} &= i g_{NN\eta}{\bar{\Psi}}_N \YY{\varphi_\eta} \Psi_N \\
    \La_{NN\sigma} &= -g_{NN\sigma}{\bar{\Psi}}_N \varphi_\sigma \Psi_N \\
    \La_{NN\rho} &= -g_{NN\rho}{\bar{\Psi}}_N \XXv{\vecvarphi_\rho\,} \Psi_N \\
    \La_{NN\omega} &= -g_{NN\omega}{\bar{\Psi}}_N \XX{\varphi_\omega}\Psi_N \\
    \La_{NN\phi} &= -g_{NN\phi}{\bar{\Psi}}_N \XXp{\varphi_\phi}\Psi_N \\
    \La_{NN\gamma} &= -e {\bar{\Psi}}_N \left( \frac{1+\tau_0}{2}
                     \gamma_\mu A^\mu +\frac{\kappa_\tau}{2m_N}
                     \sigma_{\mu\nu} \partial^\nu A^\mu \right)\Psi_N \\
    \La_{NN\gamma\varphi}&= -e\frac{g_{NN\phi}}{2m_N}{\bar{\Psi}_N} \gamma_5
                          \gamma_\mu[\tauiso \times \vecvarphi]A^\mu \;.
\end{aligned}
\end{equation}
The parameter $\chi$ controls the admixture of pseudoscalar and pseudovector
components in the corresponding Lagrangian. Its value is taken to be 0.5.
This value was obtained in our previous study of photoproduction of associated
strangeness~\cite{uso05} and has been held fixed in the study of all other 
reactions within our model. Nucleon spinors are depicted
by $\Psi$ and meson fields by $\varphi$. The magnetic moments are represented
by $\kappa$. $\La_{NN\gamma\varphi}$ generates the seagull or the contact term
diagrams. We have followed the notations of Ref.~\cite{bjo64}.

The Lagrangians for the meson vertices are
\begin{equation}\eqlab{LagrangianM}
  \begin{aligned}
   \La_{\rho\pi\pi} &= - g_{\rho\pi\pi}
             {\vecvarphi_\rho}_\mu \cdot (\vecvarphi_\pi \times \lrpart^\mu
             \vecvarphi_\pi) /2 \\
    \La_{\gamma\pi\pi} &= e \eps_{3ij} A_\mu
       (\varphi_{\pi_i} \lrpart^\mu \varphi_{\pi j}) \\
    \La_{\rho\gamma\pi} &= e \frac{g_{\rho\gamma\pi}}{m_\pi}\dop
     {\vecvarphi_\pi}{ \EPS{\partial}{A}{\partial}{\vecvarphi_\rho\,}} \\
    \La_{\omega\gamma\pi} &= e \frac{g_{\omega\gamma\pi}}{m_\pi}
                  \varphi_{\pi^0}\EPS{\partial}{A}{\partial}{\omega} \\
    \La_{\phi\gamma\pi} &= e \frac{g_{\phi\gamma\pi}}{m_\pi} \vecvarphi_{\pi^0}
                   \EPS{\partial}{A}{\partial}{\phi} \\
    \La_{\phi\gamma\eta} &= e \frac{g_{\phi\gamma\eta}}{m_\pi} \varphi_\eta
        \EPS{\partial}{A}{\partial}{\varphi_\phi} \\
    \La_{\rho\gamma\eta} &= e \frac{g_{\rho\gamma\eta}}{m_\pi} \varphi_\eta
        \EPS{\partial}{A}{\partial}{{\varphi_{rho^0}}} \\
    \La_{\rho\gamma\sigma} &= e \frac{g_{\rho\gamma\sigma}}{m_\rho}
        (\partial^\mu \varphi_{\rho^\nu} \partial_\mu A_\nu
          - \partial^\mu \varphi_{\rho^\nu} \partial_\nu A_\mu ) \\
    \La_{\rho\rho\gamma} &= 2e \big(
        A^\mu (\partial_\mu {\varphi_\rho}_\nu) \tau_0 {\varphi_\rho}^\nu
        -(\partial^\nu A^\mu) {\varphi_\rho}_\nu \tau_0 {\varphi_\rho}_\mu \\
        &\quad +(\partial^\nu A^\mu) {\varphi_\rho}_\mu \tau_0
        {\varphi_\rho}_\nu \big)\\
    \La_{\phi \K \K} &= -ig_{\phi \K \K}{\bar \vecvarphi_\K} \lrpart^\mu
                        \vecvarphi_\K) \phi_\mu\\
    \La_{\eta\K^*\K}&= -i g_{\eta\K\K^*} \vecvarphi_\K \lrpart^\mu
                       \varphi_\eta {\bar {\vecvarphi_{\K^*}}}_\mu\\
    \La_{\pi\K^*\K}& = -i g_{\pi\K\K^*} {\bar \vecvarphi_\K} \lrpart^\mu
                      \vecvarphi_\pi \cdot \tauiso {\vecvarphi_{\K^*}}_\mu \\
    \La_{\rho\pi\eta}&= -i g_{\rho\pi\eta} (\varphi_\eta \lrpart^\mu
                        \vecvarphi_\pi) {\vecvarphi_\rho}_\mu\\
    \La_{\K^*\K^0\gamma}& = \frac{g_{\K^*\K\gamma}}{m_\pi}
                  {\bar \vecvarphi_{\K^0}}
        \EPS {\partial}{A}{\partial}{\vecvarphi_{\K^*}}\\
    \La_{\K^*\K^\pm\gamma}& = \frac{g_{\K^*\K\gamma}}{m_\pi}
                  {\bar \vecvarphi_{\K^\pm}}
       \EPS {\partial}{A}{\partial}{\vecvarphi_{\K^*}} \;.
  \end{aligned}
\end{equation}
The coupling constants entering into Eqs. (A.1) and (A.2) together with
baryon magnetic moments are listed in Table III.
\begin{table}
  \caption{\tbllab{parameters} Parameters summary table}
  \begin{ruledtabular}
  \begin{tabular}{C|C|C|C}
    g_{NN\pi}            & 13.47   & g_{NN\eta}           &  0.85  \\
    g_{NN\sigma}         & 10.0    & g_{NN\rho}           & -2.2   \\
    g_{NN\omega}         & -3.0    & g_{NN\phi}           & -0.0   \\
    g_{N\Lambda \K}      & 12.0    & g_{N\Sigma K}        &  8.6   \\
    g_{N\Lambda \K^*}    & -1.7    & g_{N\Sigma K^*}      &  0.0   \\
    g_{\Sigma\sigma\rho} &-10.0    & g_{\Sigma\Lambda\rho}& 10.0   \\
    g_{\phi\K \K}        & -4.5    & g_{\rho \K \K}       & -3.0   \\
    g_{\pi \K \K^*}      & -3.26   & g_{\eta \K \K^*}     & -3.2   \\
    g_{\rho\pi\pi}       & 6.0     & g_{\rho \pi \eta}    &  0.0   \\
    g_{\rho\pi^0\gamma} & -0.12    & g_{\rho\pi^\pm\gamma}& -0.10  \\
    g_{\rho\eta\gamma}  & -0.21    & g_{\omega \eta\gamma}& -0.12  \\
    g_{\omega\pi\gamma}  & 0.32    & g_{\rho\sigma\gamma} & 12.0   \\
    g_{\phi\pi\gamma}    & 0.018   & g_{\phi\eta\gamma}   &  0.096 \\
    \kappa_p             & 1.79    & \kappa_n             & -1.91  \\
    \kappa_\Lambda       &-0.613   & \kappa_\Sigma^0      &  0.79  \\
    \kappa_{\Sigma^+}      & 1.45    & \kappa_\Sigma^-      & -0.16  \\
    \kappa_{\Sigma^0 \rightarrow \Lambda \gamma} & -1.61   &        \\
  \end{tabular}
  \end{ruledtabular}
\end{table}

For the $S_{11}$, $S_{31}$,$P_{11}$ and $P_{3,1}$ resonances the hadronic
couplings are written as
\begin{eqnarray}
\La_{\varphi NR_{1/2}} & = & -g_{\varphi NR} {\bar{\Psi}}_R [\chi{i\Gamma}
        {\varphi}+(1-\chi)\frac{1}{M}\Gamma \gamma_\mu(\partial^\mu \varphi)]
                  \Psi_N + {\rm H.c.},
\end{eqnarray}
where $M \,=\,(m_R \,\pm\,m_N)$, with upper sign for even parity and
lower sign for odd parity resonance. 
The operator $\Gamma$ is $\gamma_5$ and unity for even and odd parity
resonances, respectively.  For isovector mesons, $\varphi$ in Eq. (A.3) needs
to be replaced by $\tauiso \cdot \vecvarphi$ for isospin-$\frac{1}{2}$
resonances and by $\bf T \cdot \vecvarphi$ otherwise.

The corresponding electromagnetic couplings are
\begin{eqnarray}
\La_{\gamma NR_{1/2}} & = & -eg_1 {\bar{\Psi}_R} \frac{\Gamma}{4m_N}
                          \sigma_{\mu\nu}\Psi_N F^{\mu\nu} + H.c.,
\end{eqnarray}
where $\Psi_R$ is the resonance spinor and $F^{\mu\nu} = \partial^\mu A^\nu -
\partial^\nu \A^\mu$. The operator $\Gamma$ is 1 for the positive
parity resonance and $-i\gamma_5$ for the negative parity one.

For spin-$\frac{3}{2}$ resonances, we have used the gauge-invariant effective
Lagrangians as discussed in Refs.~\cite{kon00,pas00,pas01,pas98,luk06}. We
write here the vertex functions used by us in computation involving these
vertices.  The resonance-nucleon-pion vertex function (e.g.) is given by 
\begin{eqnarray}
\Gamma_{R_{3/2} \to N\pi}^\alpha &=& {\frac{g_1}{m_\pi}}\,
   \Big[ \gamma^\alpha (q \cdot p) -{p\!\!\!/} q^\alpha \Big]
    [(1-\chi) + \chi {p\!\!\!/} /M_p],
\end{eqnarray}
and the corresponding electromagnetic vertices are
\begin{eqnarray}
\Gamma_{R_{3/2}\to N\gamma}^{\alpha \mu} &=&   \Bigg{\{}
  (g_2 + 2 g_1) \Big[ q^\alpha p^\mu -g^{\alpha\mu} p\cdot q \Big] +
     \nonumber \\
   && g_1 \Big[ g^{\alpha\mu}{p\!\!\!/}{q\!\!\!/} - q^\alpha {p\!\!\!/}
      \gamma^\mu + \gamma^\alpha(\gamma^\mu p\cdot q - p^\mu {q\!\!\!/}) \Big]
      + \nonumber \\
   && g_3 \Big[(-q^2 g^{\alpha\mu} + q^\mu q^\alpha){p\!\!\!/} +
             (q^2 p^\mu - q^\mu p\cdot q) \gamma^\alpha \Big]
   \Bigg{\}} \nonumber \\
   &&\times \gamma_5 [(1-\chi) + \chi {p\!\!\!/} /M_p].
\end{eqnarray}
Here $p$ is the four-momentum of the resonance and $q$ is that of the meson.
Index $\alpha$ belongs to the spin-$\frac{3}{2}$ spinor and $\mu$ to photon.
Interesting property of these vertices is that the product, $p \cdot \Gamma$
= 0, where $\Gamma$ defines the vertices on the left hand side of Eqs.~(A.5)
and (A.6).  As a consequence, the spin-$\frac{1}{2}$ part of the corresponding
propagator becomes redundant as its every term is proportional to either
$p_\mu$ or $p_\nu$. Thus only spin-$\frac{3}{2}$ part of this propagator
gives rise to non-vanishing matrix elements.

\end{document}